\documentclass[useAMS,usenatbib]{mn2e}
\usepackage{graphicx}

\newcommand{\sm}{\,{\rm M}_{\odot}}
\newcommand{\kms}{\,{\rm km~s^{-1}}}
 
\begin{document} 
\pagerange{\pageref{firstpage}--\pageref{lastpage}} \pubyear{2010} 
\label{firstpage}

\title[Testing formation mechanisms of the Milky Way's thick disc]{Testing formation mechanisms of the Milky Way's thick disc with RAVE} 
 
\author [Wilson et al.]{ 
\parbox[t]{\textwidth}{ 
Michelle L.~Wilson$^{1}$\thanks{E-mail:mlw36@case.edu},  
Amina Helmi$^{2}$\thanks{E-mail:ahelmi@astro.rug.nl},  
Heather L.~Morrison$^{1}$,  
Maarten A.~Breddels$^{2}$,  
O. Bienaym\'{e}$^{3}$,  
J. Binney$^{4}$,  
J. Bland-Hawthorn$^{5}$,  
R. Campbell$^{6,7}$,  
K.C. Freeman$^{8}$,  
J.P. Fulbright$^{9}$ 
B.K. Gibson$^{10}$,  
G. Gilmore$^{11}$,  
E.K. Grebel$^{12}$,  
U. Munari$^{13}$,  
J.F. Navarro$^{14}$,  
Q.A. Parker$^{5,7}$,  
W. Reid$^{7}$, 
G. Seabroke$^{15}$,  
A. Siebert$^{3}$,  
A. Siviero$^{12,13}$,  
M. Steinmetz$^{6}$,  
M.E.K. Williams$^{6}$,  
R.F.G. Wyse$^{9}$,  
T. Zwitter$^{16}$ 
} 
\\ 
\\ 
$^{1}$Department of Astronomy, Case Western University, Cleveland, OH, 44106, USA\\ 
$^{2}$Kapteyn Astronomical Institute, P.O. Box 800, Groningen, The Netherlands\\ 
$^{3}$Universit\'{e} de Strasbourg, Observatoire Astronomique, Strasbourg, France\\ 
$^{4}$Rudolf Peierls Centre for Theoretical Physics, Oxford\\ 
$^{5}$Anglo-Australian Observatory, Sydney, Australia\\ 
$^{6}$Astrophysikalishes Institut Potsdam, Potsdam, Germany\\ 
$^{7}$Macquary University, Sydney, Australia\\ 
$^{8}$Australian National University, Canberra, Australia\\ 
$^{9}$Johns Hopkins University, Baltimore, MD, USA\\ 
$^{10}$University of Central Lancashire, Preston\\ 
$^{11}$Institute of Astronomy, Cambridge\\ 
$^{12}$Astronomisches Rechen-Institut, Zentrum f\"{u}r Astronomie der Universit\"{a}t Heidelberg, Heidelberg, Germany\\ 
$^{13}$INAF Astronomical Observatory of Padova, 36012 Asiago, Italy\\ 
$^{14}$University of Victoria, Victoria, Canada\\ 
$^{15}$Mullard Space Science Laboratory, University College London,
Holmbury St Mary, Dorking, RH5 6NT, UK\\
$^{16}$Faculty of Mathematics and Physics, University of Ljubljana, Ljubljana, Slovenia\\ 
}


\maketitle

\begin{abstract}
  We study the eccentricity distribution of a thick disc sample of
  stars observed in the Radial Velocity Experiment (RAVE) and compare
  it to that expected in four simulations of thick disc formation in
  the literature (accretion of satellites, heating of a primordial
  thin disc during a merger, radial migration, and gas-rich mergers),
  as compiled by \citet{laura}.  We find that the distribution of our
  sample is peaked at low eccentricities and falls off smoothly and
  rather steeply to high eccentricities.  This distribution is fairly
  robust to changes in distances, thin disc contamination, and the
  particular thick disc sample used.  Our results are inconsistent
  with what is expected for the pure accretion simulation, since we
  find that the dynamics of local thick disc stars implies that the
  majority must have formed \emph{in situ}. Of the remaining models
  explored, the eccentricity distribution of our stars appears to be
  most consistent with the gas-rich merger case.
\end{abstract}


\section{Introduction}

The thick disc has been a known component of the Milky Way for over 20
years \citep[]{yoshii,gilmore}. Analogous components have been
identified in external galaxies, revealing that thick discs may be
quite generic features \citep[]{burstein,kruit,yoachim}.  Most of the
thick disc formation scenarios that have been proposed fall into one
of the following four categories: accretion, heating via a minor
merger, radial migration, and intense star formation in a gas-rich
turbulent environment.

In the accretion model, satellites infall on coplanar orbits and form
the thick disc as they are disrupted and incorporated into the main
galaxy; in this case, the thick disc would thus consist of stars
originating largely \citep[more than 70 per cent in the simulations
of][]{abadi} in several disrupted satellites.

The most often discussed scenario for the formation of a thick disc is
the accretion of a massive satellite by a pre-existing disc galaxy,
which is thus heated dynamically. The resulting thick disc is mainly
made from stars that originated in the primary galaxy's primordial
thin disc rather than the merging satellite. For example in the
simulations by \citet{alvaro}, 10 to 20 per cent of the stars in the
remnant thick disc at the ``Solar'' radius are accreted.

During a turbulent gas-rich phase, stars may also form in a thick
disc. This can happen in massive clumps in a rotationally supported
component, as in the simulations of \citet{bournaud}, but also during
gas-rich mergers as shown by \cite{brook}. In the latter case, the resulting
thick disc stars would have been formed \emph{in situ} and with relatively hot
kinematics. 

Merging events of any variety might be unnecessary to explain the
phenomena, however.  Stars may migrate radially from the inner parts
of a galaxy to the outer regions due to resonant interactions with
spiral arms \citep{sellwood} and a bar \citep{minchev}. Although the
migration process itself does not heat the disc, a greater vertical
velocity in the higher surface brightness central regions results in
larger heights above the plane being reached in the outer regions
where the surface brightness is lower \citep{kregel}.  Migration can
thus result in the formation of a thick disc from thin disc stars
without any external stimulus \citep{roskar,schoenrich}.

Recently, \citet{laura} suggested a dynamical test that could
differentiate between the formation models and be applied to data to
disentangle which is the most likely to have occurred in our galaxy:
they propose using the eccentricities of thick disc stars as
a discriminant, as populations formed \emph{in situ} are likely to move on
low eccentricity orbits, while those accreted can have any
eccentricity, but will typically be biased towards higher values.  As
kinematic data for thick disc stars become available from surveys
such as the Radial Velocity Experiment \citep[RAVE][]{steinmetz2006},
SEGUE \citep{yanny2009}, and eventually
{\it Gaia}, we can apply these tests to see what they can reveal
about thick disc formation.

In this paper, we investigate and constrain the dynamics of a sample
of Milky Way thick disc stars and compare the resulting eccentricity
distribution to what is expected for the above four models.  We use
data from the RAVE survey and distances calculated in the manner of
\citet{maarten} but with modifications described in Section
\ref{sec:data}. In Section \ref{sec:results} we derive the
eccentricity distribution of the stars and in Section \ref{sec:theory} we then
examine this in light of the accretion, heating, merger, and migration
simulations discussed in \citet{laura}.

\section{Data}
\label{sec:data}

\subsection{The RAVE survey}
\label{ssec:rave}

RAVE\footnote{We use the 30 August 2008 internal data release, which
  consists of 135,338 stars.} measures radial velocities and stellar
atmospheric parameters from spectra using the 6dF multi-object
spectrometer on the Anglo-Australian Observatory's 1.2 m UK Schmidt
Telescope.  The survey looks in the Ca-triplet region (8410-8795 \AA),
has a resolution of $\sim 7500$, and is magnitude limited.  The targets chosen
are southern hemisphere stars taken from the Tycho-2, SuperCOSMOS and DENIS
surveys with I-band magnitudes between 9 and 13.  The average internal
errors in radial velocity (RV) are $\sim 2 \kms$, and the approximate
RV offset between RAVE and the literature is smaller than $\sim 1
\kms$.  The catalogue also includes 2MASS photometry and proper motions
from Starnet 2.0, Tycho-2, SuperCOSMOS, and UCAC2.  For more
information about RAVE, see \citet{zwitter}.

\subsection{Distances} 
\label{ssec:distances} 
 
 
Distances were calculated along the lines of \citet{maarten}.  We 
briefly sketch the method here and refer the reader to \citet{maarten} for 
more details. 
 
The stars are fit using the $Y^2$ (Yonsei-Yale) isochrones \citep{y2} 
and their measured characteristics.  The measured quantities of the 
stars ($T_{\rm eff}$, log(g), [M/H], $J$, and $J-K_{s}$) from the RAVE 
pipeline \citep{zwitter} and 2MASS are used to minimise the $\chi^2$ 
statistic to find the closest model star for each observed star. 
Then, the errors of the observed quantities, which are assumed to be 
Gaussian, are utilised in a Monte Carlo simulation, from which the 
absolute magnitude and its error are determined from the resulting 
probability distribution function.  Stars for which none of the 
isochrones provide adequate matches are discarded.  Since the $Y^2$ 
isochrones do not extend past the RGB tip, clump stars may result in 
poor fits; in addition we have explicitly removed the clump stars that 
should remain in the data set (see Section \ref{ssec:sample}). 
 
\begin{center} 
\begin{figure} 
\includegraphics[width=84mm]{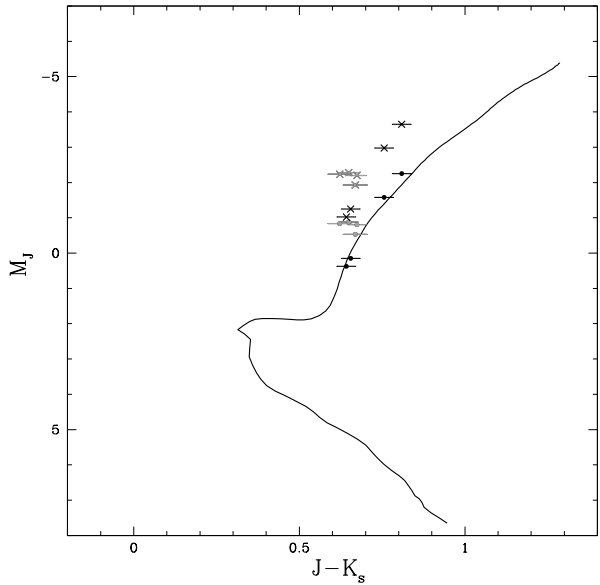} 
\caption{The theoretical solar metallicity, 4 Gyr isochrone with the 
  CMD of M67 giants.  Points with crosses are the M67 stars 
  transformed to absolute magnitudes using the \citet{maarten} 
  distance calculated after excluding the clump stars of 1.48 kpc; filled points use the distance of 0.78 kpc 
    calculated assuming an age of 4 Gyr.  Gray points are stars 
    identified as belonging to the red clump.} 
\label{fig:M67} 
\end{figure} 
\end{center}

The effectiveness of this distance method was examined in detail in 
\citet{maarten}, but the main points will be mentioned here. 
Comparison to main sequence stars observed by HIPPARCOS showed good 
agreement in the measured parallaxes. To test the distances for 
giants, \citet{maarten} used the members of M67, an old open cluster. 
The mean of the calculated distances was $1.48 \pm 0.36$ kpc, compared 
to 0.78 kpc from \citet{vandenberg}.  This discrepancy is related to 
the fact that when age is left as a variable, stars on the giant 
branch are sometimes fit better by (unrealistically) younger 
isochrones.  However, when using the 4 Gyr isochrones, which 
corresponds to the accepted age for the cluster, the distances agreed 
with those in the literature within error (see Figure \ref{fig:M67}). 
 
This has motivated us to calculate distances for all stars setting the 
age at 10 Gyr, which is the characteristic age of the thick disc 
\citep{edvardsson}.  This implies that our method will assign 
incorrect distances to younger thin disc stars, but since younger 
giant branches are brighter than older ones, this assumption results in the 
assigned distances to younger stars being slightly smaller than they 
actually are. As discussed in Sec.~\ref{ssec:sample}, we will be 
selecting stars with 1 kpc~$\le |z| \le$~3 kpc, so as to isolate a 
thick disc sample. The consequence of using the ``wrong'' isochrone 
for young stars is to reduce their $|z|$ and to move many from lower 
$|z|$ out of our sample rather than scattering thin disc stars up into 
it.  Even so, we investigated what effect that would have on the 
distances and the level of thin disc contamination more quantitatively 
in Section \ref{sssec:thindisc}.

\begin{center} 
\begin{figure} 
\includegraphics[width=84mm]{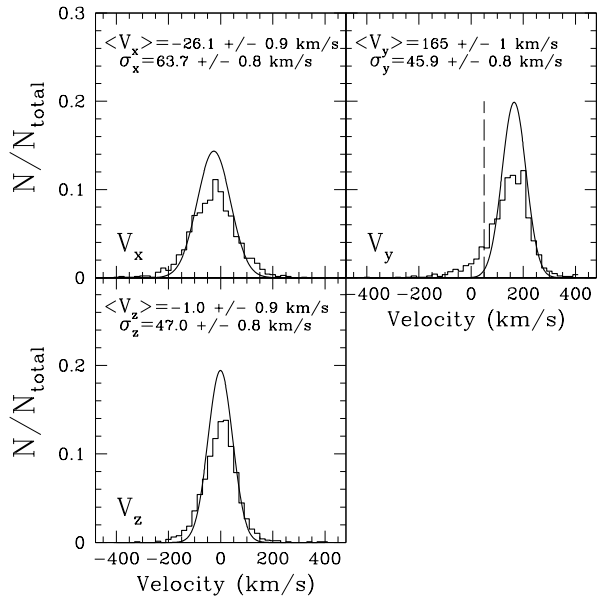} 
\caption{Velocity distributions of our thick disc sample (1273 stars) 
computed in a right-handed reference frame at rest with respect to the 
Galactic centre.  Mean velocity values and dispersions, calculated 
using robust estimation, for each component are given in each panel. 
Stars to the left of the dashed line in $V_{y}$ were discarded when 
calculating the means and dispersions.} 
\label{fig:velsbreddels} 
\end{figure} 
\end{center} 
 
\subsection{Sample selection} 
\label{ssec:sample} 
 
We first cleaned the data set by discarding any stars with distance 
errors $>$ 40 per cent, proper motion errors in either RA or DEC $>$ 10 
mas/yr, or radial velocity errors $> 5 \kms$.  Once the age was set at 
10 Gyr, the clump stars were not fit well by any of the isochrones, so 
most of them should have been discarded on that basis.  To ensure all 
clump stars were indeed removed, however, we threw out stars with 
${\rm log(g)} > 1.5$ and $J-K_s < 0.75$.

In order to isolate a sample of thick disc stars, we chose the
remaining stars with $|z|$ between 1 and 3 thick disc scale-heights,
which corresponds to the range 1 -- 3 kpc \citep[]{veltz08}.  The
decision to make our thick disc selection based only on $|z|$ rather
than including a metallicity criterion was motivated partly by
uncertainties in the RAVE's metallicity pipeline\footnote{For
  completeness, we have tested that our results do not change
  appreciably when we focus only on the subset of stars with high S/N spectra which
  according to their [M/H] belong to the thick disc.}, but also because
this mimics more closely the selection by \citet{laura}.  We further
clipped our sample, discarding all stars with $V_{y} < 50 \kms$ to
minimise contamination from the halo, and including only stars within
a heliocentric cylinder with a radius of 3 kpc so the data was in a
form best suited for comparison with the eccentricity distributions of
models as illustrated by \citet{laura}.  The final sample consisted of
1273 stars.  The velocity distributions are given in Figure
\ref{fig:velsbreddels}.  The velocity dispersions are generally
consistent with the literature values, especially in $V_{x}$, while
those in $V_{y}$ and $V_{z}$ are slightly higher by approximately $6-8
\kms$ \citep{sandg}.
 
\section{Results} 
\label{sec:results}

\subsection{Eccentricity distribution} 
\label{sec:eccentricity} 
 
To calculate the eccentricities of the RAVE stars in our thick disc 
sample we integrated their orbits in a Galactic potential. This 
consisted of a \citet{miyamoto75} disc, a Hernquist (1990a) bulge and 
a spherical logarithmic halo.  In this model, the characteristic 
parameters used were $M_{\rm disc} = 8.0 \times 10^{10}, M_{\rm bulge} = 2.5 
\times 10^{10}, v_{\rm halo}^2 = 27000.0, a = 6.5, b = 0.26, c = 0.7$, and 
$d = 12.0$, with masses in $ \sm $, velocities in $\kms$, and lengths in 
kpc, which produce a circular velocity of $\sim 220 \kms$ at 8 kpc 
from the Galactic centre. The eccentricities of the stars were defined 
as $(r_{\rm apo}-r_{\rm peri})/(r_{\rm apo}+r_{\rm peri})$, where
$r_{\rm apo}$ 
($r_{\rm peri}$) is the maximum (minimum) distance reached by the star in 
its orbit. Figure \ref{fig:distwig20} shows the eccentricity 
distribution obtained for our sample. The main features of this 
distribution are an asymmetric peak at a fairly low eccentricity value 
with a relatively (though not entirely) smooth falloff towards higher 
eccentricities.  This peak at low eccentricity suggests that the thick 
disc was formed primarily \emph{in situ} (see Section 
\ref{ssec:comparemodels}). 
 
\subsection{Robustness of the eccentricity distribution} 
\label{ssec:robustness} 
 
\subsubsection{Systematic errors} 
\label{sssec:systematic} 
 
We expect that the largest contribution to systematic errors would be 
due to the distances.  Distance overestimation would result in larger 
heliocentric velocities, because the observed proper motions would be placed 
at larger distances.  Thus, the distance overestimation should result 
in a peak at a higher eccentricity than more accurate distances would 
reveal, not a peak at lower eccentricities, suggesting that the 
calculated distribution would not likely be a result of distance 
overestimation.  On the other hand, distance underestimation should 
result in smaller heliocentric velocities and more circular orbits, and thus could 
cause an artificially strong peak at low eccentricity. 
 
\begin{center} 
\begin{figure} 
\includegraphics[width=84mm]{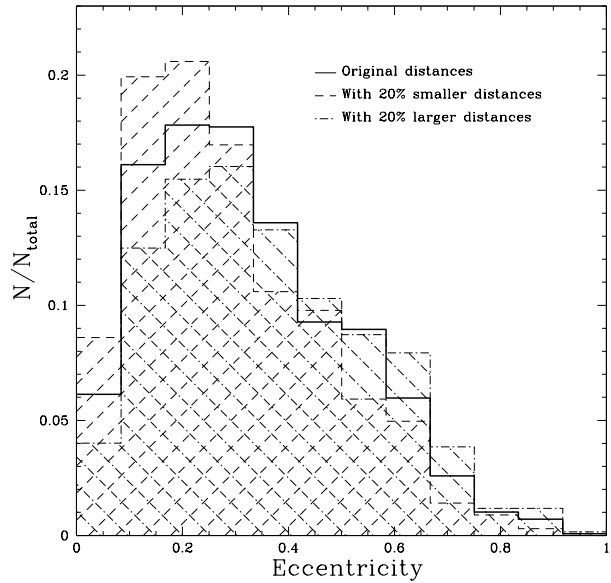} 
\caption{Eccentricity distribution of the thick disc sample using the 
  original distances (1273 stars, solid histogram) and distances that 
  are 20 percent smaller (1350 stars, dashed histogram) and 
  larger (1204 stars, dashed-dotted histogram).  The differing 
  numbers of stars are a result of discarding stars with low 
  rotational velocity, since this number depends on what distance is 
  assumed.} 
\label{fig:distwig20} 
\end{figure} 
\end{center}

To explore more quantitatively how errors in distance could affect the 
final eccentricity distribution, we calculated eccentricities for the 
thick disc sample using distances that were larger and smaller by 20 
percent.  This value was chosen because nearly all the stars in the 
sample had distance errors smaller than 20 per cent, with a peak at 10 
percent, making this figure conservative.  As Figure 
\ref{fig:distwig20} shows, the distribution with smaller distances has 
a slightly higher peak at a slightly smaller eccentricity than the 
original distribution, and that with larger distances has a slightly 
lower peak at a slightly larger eccentricity, as expected.  However, 
both distributions remain reasonably close to the original one, which 
suggests that a systematic distance error of 20 percent would not 
result in a substantial change in the thick disc's eccentricity 
distribution. We have found that even with up to 40 percent larger 
distances, the peak in eccentricity remains below 0.4 and the 
generally triangular shape does not change. 

We have also tested how our proper motion errors affect the
eccentricity distribution. We found that the main effect is to
slightly lower the amplitude of the peak at eccentricity $\sim 0.2$
and to marginally increase the number of stars with high
eccentricity. Therefore we conclude that the shape of the eccentricity
distribution is generally robust to the estimated uncertainties in our
observables.

\subsubsection{Investigation of possible thin disc contamination} 
\label{sssec:thindisc} 
 
Since the thin disc characteristically has stars on fairly circular 
orbits, there is a possibility that the peak at low eccentricity is 
caused partially by thin disc contamination in our sample. 
 
To give a quantitative estimate of this contamination, we formulated a 
simple model.  Using the Padova isochrones and the \citet{chabrier} 
initial mass function (IMF), we generated two samples of stars.  A 
thin disc population was created from a solar metallicity, 5 Gyr 
isochrone, and a thick disc population was created from an isochrone 
with [Fe/H] = $-0.6$~dex and an age of 10 Gyr.  The relative fraction 
of thin to thick disc stars $f_{\rm thin2thick}$ was calculated using the 
expression 
\[ f_{\rm thin2thick}= f_{\rm norm}*e^{-|z|/z_{\rm thin}+|z|/z_{\rm thick}},\] 
where $f_{\rm norm}$ is the relative fraction at the Sun, and $z_{\rm thin}$ 
and $z_{\rm thick}$ the thin and thick disc scale heights respectively. 
We assumed $z_{\rm thin}= 225$ pc and $z_{\rm thick} = 1048$ pc, as estimated 
by \citet{veltz08} for the RAVE sample.  The local normalisation 
$f_{\rm norm}$ was calculated using the \citet{veltz08} ratio of thin to 
thick disc dwarfs and the Padova isochrones to determine the fraction of 
dwarfs for the thin and thick disc samples.  We assigned $z$ from 1050 
pc to 2950 pc in increments of 100 pc and selected the model stars 
that would have been observed by RAVE $(9 < I < 13)$.  To synthesise 
the effect of calculating distances assuming an age of 10 Gyr on the 
younger thin disc population, we found the absolute magnitude on the 
giant branch that would have been assigned to each thin disc star 
using a solar metallicity, 10 Gyr isochrone based on their effective 
temperatures.  New distances were calculated for the stars with the 
new absolute magnitudes and the original apparent magnitudes and then 
re-binned based on these new distances. 
 
Ratios of the number of observed thin disc stars (both using the 5 and 
10 Gyr ages) to observed thick disc stars were then computed for each 
bin in distance and overall.  In the lowest $z$ bin, the thin disc 
contamination was as large as 30 percent, but it dropped to about 10 
percent for the 10 Gyr thin disc by $z = 1250$ pc and by $z=1350$ pc 
for the 5 Gyr thin disc.  Since the effect of calculating the 
distances of younger stars with 10 Gyr isochrones is to underestimate 
their distances, the thin disc contamination was lower at most $z$ 
intervals for the 10 Gyr thin disc ratios, since moving the stars down 
in $z$ took many out of the $z$ range considered.  The overall  
thin disc contamination for the 5 and 10 Gyr thin discs were, 
respectively, 4.7 and 3.1 percent. 
 
\begin{center} 
\begin{figure} 
\includegraphics[width=84mm]{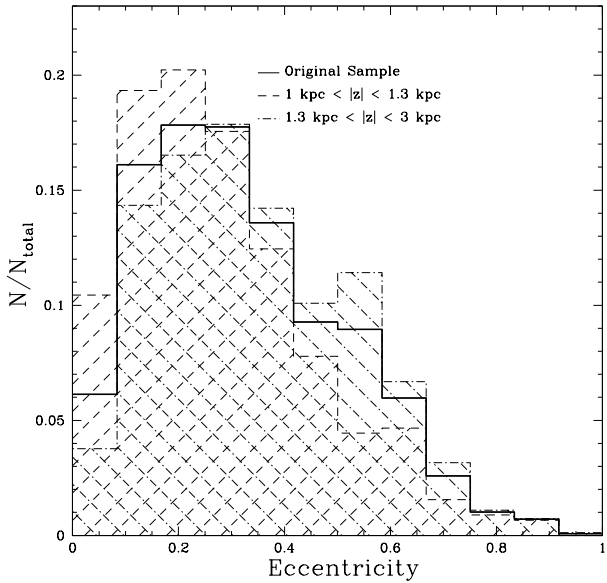} 
\caption{Eccentricity distributions of the original sample (solid),
  the subsample of stars with 1 kpc $< |z| <$ 1.3 kpc (dashed
  histogram, 450 stars), and the subsample with 1.3 kpc $< |z| <$ 3
  kpc (dashed-dotted histogram, 823 stars).}
\label{fig:split13} 
\end{figure} 
\end{center} 
 
Since those ratios suggest that there could be significant 
contamination in the portion of our sample that is closest to the 
plane of the disc, we took our thick disc sample and divided it into a 
low $|z|$ and a higher $|z|$ portions to see how robust the shape of 
the eccentricity distribution was to thin disc contamination.  Two 
separate trials were performed.  In the first, the division was placed 
at $|z|$ = 1.3 kpc, because the simple model suggests that the 
contamination levels have dropped below 10 percent by that height when 
the stars all were assigned 10 Gyr ages, as is the case for the data 
set.  The resulting eccentricity distributions for each subsection 
were then calculated and are plotted in Figure \ref{fig:split13}.  The 
distribution of the lower $|z|$ portion is strongly peaked at low 
eccentricity with fewer stars at higher eccentricities, which would be 
expected for a sample contaminated by thin disc stars, which have 
predominantly circular orbits.  The higher $|z|$ sample retained the 
roughly triangular shape of the original sample, however.  The peak 
did shift slightly to higher eccentricity, but not significantly. 
 
We also performed another trial and cut the sample at $|z|$ = 1.5 kpc.
The resulting eccentricity distributions of the two subsamples were
similar to those of the first trial.  The distribution of stars in the
higher $|z|$ subsample was lumpier, since there were fewer stars in
it, and a slightly increased amount of higher eccentricity stars was
noticeable. These trends are natural consequences of increased
contamination by halo stars.  Overall, the higher $|z|$ distribution
retained the general properties of the original distribution.  Thus we
conclude that the thin disc contamination is not likely to have a
large effect on the overall shape of the eccentricity distribution.
 
\subsubsection{Comparison of the original eccentricity distribution with those using different thick disc samples} 
\label{sssec:compare} 
 
\begin{center} 
\begin{figure} 
\includegraphics[width=84mm]{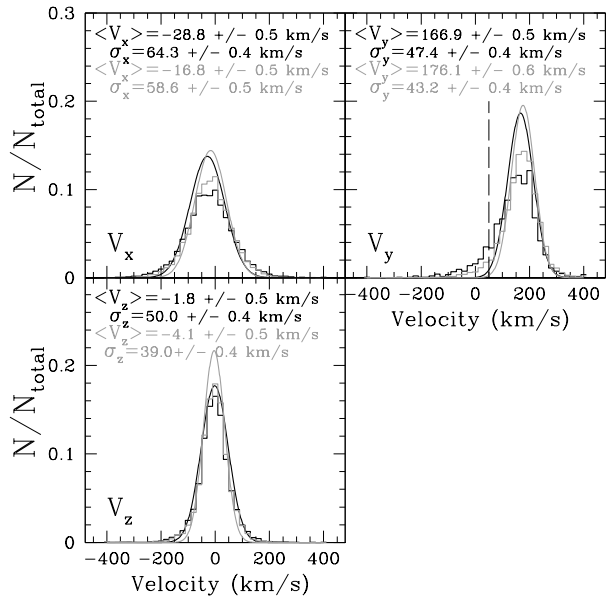} 
\caption{Velocity distributions of the thick disc sample using the 
  distances of \citet{zwitter-dist} (black) and the clump 
  thick disc sample (gray).  The mean velocity values and dispersions 
  (calculated using robust estimation) after removing stars with 
  $V_y < 50\kms$ are given in each panel.  The Gaussians plotted have 
  the same mean and dispersion as the data.} 
\label{fig:velszwitter} 
\end{figure} 
\end{center}

Since the distances are a key component in calculating the
eccentricity distribution, we also used another set of distances.
\citet{zwitter-dist} calculated distances for a RAVE data set (based
on a later internal data release consisting of 332,747 stars) by means
of a photometric parallax method based on that of \citet{maarten} but
using the $Y^2$, Dartmouth \citep{dotter}, and Padova isochrones
uniformly, rather than logarithmically spaced in time and, when
picking a best match model star, weighting the model star picked based
on mass. No age constraint to tailor the distances to a thick disc
sample was applied in this set of distances.  Using the same cleaning
and thick disc selection criteria as before, we selected a sample from
this catalogue (that based on the $Y^2$ isochrones) and calculated its
eccentricity distribution.  This thick disc sample consists of 6173
stars.  See Figure \ref{fig:velszwitter} for the velocity
distributions.

In order to have a more independent check on the eccentricity 
distribution, since the above two methods of calculating distances are 
quite similar, we selected a thick disc sample of clump stars from 
RAVE.  To choose this sample, we used the same cleaning criteria on 
the full data release but did not impose the restrictions aimed at 
eliminating the clump stars.  We identified the clump by colour and gravity, 
taking it to have $0.6 < J-K_s < 0.7$ and 1.5 $<$ log(g) $<$ 2.7.  Then, we 
calculated the distances using the apparent magnitudes and assuming an 
absolute clump magnitude of $M_K = -1.61$, 
as in \citet{clumpmag}.  Finally, we imposed the same restrictions in 
$|z|$, volume, and velocity as before and calculated the eccentricity 
distribution.  The resulting sample includes 3573 stars, and the 
velocity distributions are given in Figure \ref{fig:velszwitter}.

%
\begin{figure} 
\includegraphics[width=85mm]{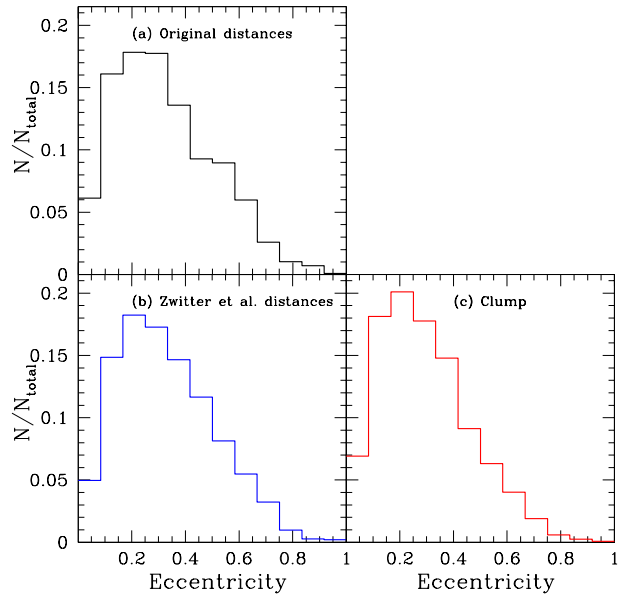} 
\caption{Eccentricity distributions of thick disc samples  
  using a) the original distances, b) the Zwitter et al. distances, and c) the clump sample.} 
\label{fig:all3separate} 
\end{figure} 
 
The velocity distributions in Figure~\ref{fig:velszwitter} show good
agreement with those based on the \citet{maarten} distances.  As
expected, the velocity dispersions for the red clump sample are
slightly lower, since this sample should be devoid of halo
contamination (as red clump stars are only present in young or
intermediate age populations). Figure \ref{fig:all3separate} shows
that also the thick disc eccentricity distributions agree well with
each other.  All are strongly peaked at low eccentricities, have a
generally triangular shape, and do not have a secondary peak at high
eccentricity.  The sample based on \citet{maarten} falls off at high
eccentricities less smoothly than the other two, but that could be due
to statistical fluctuations in this much smaller data set.  On the
other hand, the clump sample has its peak at a lower eccentricity and
contains a smaller number of stars at higher eccentricities as
expected since this region is populated mainly by halo stars which are
only present in the red giant samples.  So while the three
distributions are not identical, they agree fairly closely.  Results
of the application of KS tests show that all three are consistent with
being drawn from the same distribution. This leads us to conclude that
the eccentricity distribution is reasonably robust to the exact thick
disc sample selection.

\begin{center} 
\begin{figure} 
\includegraphics[width=84mm]{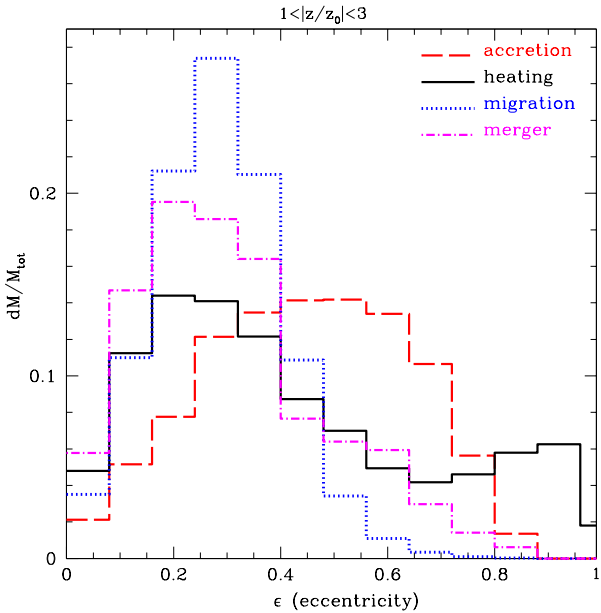} 
\caption{Comparison of the eccentricity distributions of each thick 
  disc formation model for stars in the range 1--3 (thick disc) 
  scale-heights and cylindrical distance $2 < R/R_d < 3$. } 
\label{fig:e_all} 
\end{figure} 
\end{center} 
 
\section{Comparison with theoretical models} 
\label{sec:theory} 
 
\subsection{Discussion of the models} 
\label{ssec:models} 
 
We now compare our calculated eccentricity distribution to those 
computed by \citet{laura} from the simulations of \citet{abadi}, 
\citet{alvaro-08}, \citet{roskar}, and \citet{brook}.  These simulations 
have been discussed in the literature, so we will only briefly 
describe them here.  For details, we refer the reader to the 
literature and to Table 1 of \citet{laura}, which summarises their 
main parameters. 
 
\citet{abadi} demonstrate the formation of the thick disc through 
accretion of satellites during the hierarchical formation of a Milky 
Way-like galaxy in the $\Lambda$ cold dark matter paradigm and using 
cosmological \emph{N}-body/smoothed particle hydrodynamics (SPH) 
simulations. 
 
In the heating scenario simulation of \citet{alvaro}, a satellite 
merges with a primary thin disc (of comparable mass) on a prograde 
orbit inclined 30 degrees.  The subsequent formation of a new thin disc 
from cooling gas and any changes that might cause in the thick disc 
were not modelled in this simulation.  
 
The simulation of \citet{roskar} models the formation of a galactic 
disc by starting with a dark matter halo and a hot gas halo which over 
10 Gyrs cools and forms stars in a disc which migrate as transient 
spiral structure forms. 
 
\citet{brook} model the formation of a disc in a semi-cosmological 
\emph{N}-body/SPH simulation which incorporates gas heating and 
cooling, star formation, feedback, and chemical enrichment. Their 
thick disc develops during gas rich mergers.

\citet{laura} found that the location of the peak and the shape of the 
eccentricity distributions are driven by what kind of stars make up 
the thick disc in each model, as shown in Figure \ref{fig:e_all}.  In the 
accretion scenario, the thick disc is composed mostly of accreted 
material; since satellites tend to be on radial orbits, the resulting 
eccentricity distribution is relatively broad and peaks in the mid to 
high eccentricities.  The other three scenarios involve a thick disc 
being composed primarily from stars formed \emph{in situ}.  Such stars 
have more circular orbits, which produce a peak at low values of the 
eccentricity.  The heating and the gas-rich merger simulations have a 
contribution of accreted stars to the thick disc, which show up in the 
eccentricity distribution as a distinct second but much smaller peak 
at high eccentricity in the heating scenario and a lumpiness at mid to 
high eccentricities in the merger case.

\subsection{Comparison with the models} 
\label{ssec:comparemodels} 
  
Comparison of Figures \ref{fig:all3separate} and \ref{fig:e_all} shows 
that the eccentricity distribution of our sample does not support the 
accretion model.  The triangular shape with a peak at low 
eccentricities of our distribution does not resemble the broad mound 
peaked at middling eccentricity that characterises the accretion 
scenario.  The eccentricity distribution of our thick disc sample with 
its prominent peak at low eccentricities is consistent with the 
distribution displayed by the heating, migration, and merger 
scenarios.  Our distribution does not exhibit a secondary peak at high 
eccentricities like that evident in the heating model due to the 
accreted satellite.  However, the location of this secondary peak 
depends on the initial orbital configuration of the accreted 
satellite. If the initial conditions were changed to produce a 
secondary peak at middling eccentricities this could probably be 
obscured in the data by the larger fraction of \emph{in situ} stars. 
 
While the distributions from the migration and merger models are not 
identical to the distribution of our sample, they do show some 
resemblance and have no features that seem to count against either 
being possible.  The gas-rich merger's distribution even displays an 
asymmetric peak like our distribution, making this 
scenario the most consistent with our data.  The migration 
distribution exhibits a symmetry about the peak (is more 
Gaussian-like) until it gets down into the high eccentricity tail that 
is not displayed in our distribution, making the migration scenario 
somewhat less consistent with our data. 
 
There are several reasons for the models not being a perfect match to 
the data.  Random and systematic errors in the data, while not 
significantly affecting the overall characteristics of the 
distribution, can alter the exact shape, especially at high 
eccentricities.  Also, since the simulations did not attempt to 
duplicate the Milky Way exactly, they might not be similar enough to 
the Galaxy to precisely mimic the eccentricity distribution even if 
the formation mechanism is correct.  It is also plausible that the 
thick disc was formed by a combination of processes.  Radial migration 
has been shown to be dynamically possible in a galaxy such as our own; 
mergers, including gas-rich, are known to occur.  Thus, the 
nature of the measured eccentricity distribution could be indicating 
that both radial migration and gas-rich mergers contributed to its 
formation. 
 
One possible way to investigate which model is the most likely, would 
be to compute the eccentricity distributions of the models and the 
data for different locations along the Galactic disc, as one may 
expect the contribution of accreted populations to change with 
distance and become more dominant in the Galaxy's 
outskirts. Undoubtedly, data from {\it Gaia} when it becomes available 
will aid in clarifying how exactly the Galactic thick disc formed. 
 
\section{Summary and Conclusions} 
\label{sec:summary} 
 
We have isolated a sample of thick disc stars from RAVE survey data, 
calculated its eccentricity distribution, and determined that the 
distribution is fairly robust to changes in distances, thin disc 
contamination, and the specific thick disc sample used.  Our 
eccentricity distribution is fairly triangular in shape, depicting a 
dominant peak at low eccentricity and a relatively smooth falloff at 
high values.  
 
We have compared this finding with the eccentricity distributions in
\citet{laura} presented for simulated thick discs formed via
accretion, heating via a minor merger, radial migration, and gas-rich
mergers. The broad peak at moderately high eccentricities of the
accretion model is not consistent with the relatively narrow peak at
low eccentricity displayed by our sample.  This indicates that the
Galactic thick disc formed predominantly \emph{in situ}.  A lack of a
distinguishable secondary peak at high eccentricity further suggests
that if any of the thick disc was accreted, its direct contribution of
stars in the Solar neighbourhood was minimal. The gas-rich mergers
simulation, and to a somewhat lesser extent, the radial migration model,
are consistent with the distribution of our sample. This suggests that
these formation mechanisms could have had some role in the formation
of the Milky Way's thick disc.

\section*{Acknowledgements} 
 
We are indebted to Laura V.~Sales for producing Figure~7 of this 
Paper.  AH acknowledges financial support from the European Research 
Council under ERC-Starting Grant GALACTICA-240271. HLM has been 
supported by the NSF through grant AST-0098435. 
 
Funding for RAVE has been provided by the Anglo-Australian 
Observatory, the Astrophysical Institute Potsdam, the Australian 
Research Council, the German Research foundation, the National 
Institute for Astrophysics at Padova, The Johns Hopkins University, 
the Netherlands Research School for Astronomy, the Natural Sciences 
and Engineering Research Council of Canada, the Slovenian Research 
Agency, the Swiss National Science Foundation, the National Science 
Foundation of the USA (AST-0508996), the Netherlands Organisation for 
Scientific Research, the Particle Physics and Astronomy Research 
Council of the UK, Opticon, Strasbourg Observatory, and the 
Universities of Basel, Cambridge, and Groningen. The RAVE Web site is 
at www.rave-survey.org.

 

 

\bsp 
 
\label{lastpage} 
 

\begin{thebibliography}{99} 
 
\bibitem[\protect\citeauthoryear{Abadi et al.}{2003}]{abadi} Abadi M.~G., Navarro J.~F., Steinmetz M., Eke V.~R., 2003, ApJ, 597, 21 
 
\bibitem[\protect\citeauthoryear{Alves  
\& Sarajedini}{1999}]{clumpmag} Alves D.~R., Sarajedini A., 1999, ApJ, 511, 225 
 
\bibitem[\protect\citeauthoryear{Bournaud et 
    al.}{2009}]{bournaud} Bournaud, F., Elmegreen, B.~G., 
  \& Martig, M.\ 2009, ApJ, 707, L1 
 
 
 
\bibitem[\protect\citeauthoryear{Breddels et  
al.}{2010}]{maarten} Breddels M.~A., et al., 2010, A\&A, 511, A90 
 
\bibitem[\protect\citeauthoryear{Brook et al.}{2004}]{brook} Brook C.~B., Kawata D., Gibson B.~K., Freeman K.~C., 2004, ApJ, 612, 894 
 
\bibitem[\protect\citeauthoryear{Burstein}{1979}]{burstein} Burstein D., 1979, ApJ, 234, 829 
 
\bibitem[\protect\citeauthoryear{Chabrier}{2001}]{chabrier} Chabrier G., 2001, ApJ, 554, 1274 
 
\bibitem[\protect\citeauthoryear{Demarque et  
al.}{2004}]{y2} Demarque P., Woo J.-H., Kim Y.-C., Yi  
S.~K., 2004, ApJS, 155, 667  
 
\bibitem[\protect\citeauthoryear{Dotter et al.}{2008}]{dotter}  
Dotter A., Chaboyer B., Jevremovi{\'c} D., Kostov V., Baron E., Ferguson  
J.~W., 2008, ApJS, 178, 89 
 
\bibitem[\protect\citeauthoryear{Edvardsson et  
al.}{1993}]{edvardsson} Edvardsson B., Andersen J., Gustafsson B., Lambert D.~L., Nissen P.~E., Tomkin J., 1993, A\&A, 275, 101 
 
\bibitem[\protect\citeauthoryear{Gilmore \& Reid}{1983}]{gilmore} Gilmore G., Reid N., 1983, MNRAS, 202, 1025 
 
\bibitem[\protect\citeauthoryear{Hernquist}{1990}]{hern90}  
Hernquist L., 1990, ApJ, 356, 359  
 
\bibitem[\protect\citeauthoryear{Kregel, van der Kruit, \& Freeman}{2005}]{kregel} Kregel M., van der Kruit P.~C., Freeman K.~C., 2005, MNRAS, 358, 503 
 
\bibitem[\protect\citeauthoryear{Marigo et  
al.}{2008}]{marigo} Marigo P., Girardi L., Bressan A., Groenewegen M.~A.~T., Silva L., Granato G.~L., 2008, A\&A, 482, 883  
 
\bibitem[\protect\citeauthoryear{Minchev \& Famaey}{2009}]{minchev} Minchev, I., \& Famaey, B.\ 2009, arXiv:0911.1794  
 
\bibitem[\protect\citeauthoryear{Miyamoto \& Nagai}{1975}]{miyamoto75} Miyamoto M., Nagai R., 1975, PASJ, 27, 533  
 
\bibitem[\protect\citeauthoryear{Ro{\v s}kar et al.}{2008}]{roskar} Ro{\v s}kar R., Debattista V.~P., Stinson G.~S., Quinn T.~R., Kaufmann T., Wadsley J., 2008, ApJ, 675, L65 
 
\bibitem[\protect\citeauthoryear{Sales et al.}{2009}]{laura}  
Sales L.~V., et al., 2009, MNRAS, L336  
 
\bibitem[\protect\citeauthoryear{Sch{\"o}nrich  
\& Binney}{2009}]{schoenrich} Sch{\"o}nrich R., Binney J., 2009, MNRAS, 399, 1145 
 
\bibitem[\protect\citeauthoryear{Sellwood \& Binney}{2002}]{sellwood} Sellwood J.~A., Binney J.~J., 2002, MNRAS, 336, 785 
 
\bibitem[\protect\citeauthoryear{Sparke \& Gallagher}{2006}]{sandg} Sparke L.~S., Gallagher J.~S., III 2006, Galaxies in the Universe - 2nd Edition, Cambridge University Press   

\bibitem[Steinmetz et al.(2006)]{steinmetz2006} Steinmetz M., et al.,
  2006, AJ, 132, 1645
 
\bibitem[\protect\citeauthoryear{van der Kruit  
\& Searle}{1981}]{kruit} van der Kruit P.~C., Searle L., 1981, A\&A, 95, 105  
 
\bibitem[\protect\citeauthoryear{VandenBerg et  
al.}{2007}]{vandenberg} VandenBerg D.~A., Gustafsson B.,  
Edvardsson B., Eriksson K., Ferguson J., 2007, ApJ, 666, L105   
 
\bibitem[\protect\citeauthoryear{Veltz et  
al.}{2008}]{veltz08} Veltz L., et al., 2008, A\&A, 480, 753 
 
\bibitem[\protect\citeauthoryear{Villalobos \& Helmi}{2008}]{alvaro-08} Villalobos {\'A}., Helmi A., 2008, MNRAS, 391, 1806 
 
\bibitem[\protect\citeauthoryear{Villalobos \& Helmi}{2009}]{alvaro} Villalobos {\'A}., Helmi A., 2009, MNRAS, 399, 166 
 
\bibitem[Yanny et al.(2009)]{yanny2009} Yanny B., et al., 2009, AJ,
  137, 4377

\bibitem[\protect\citeauthoryear{Yoachim \& Dalcanton}{2006}]{yoachim} Yoachim P., Dalcanton J.~J., 2006, AJ, 131, 226 
 
\bibitem[\protect\citeauthoryear{Yoshii}{1982}]{yoshii} Yoshii Y., 1982, PASJ, 34, 365  
 
\bibitem[\protect\citeauthoryear{Zwitter et al.}{2008}]{zwitter} Zwitter T., Siebert A., Munari U., et al., 2008, AJ, 136, 421 
 
\bibitem[\protect\citeauthoryear{Zwitter et al.}{2010}]{zwitter-dist}
  Zwitter T., et al., 2010, arXiv:1007.4411
  
 
 
\end{thebibliography}
\end{document}